\documentclass{article}
\usepackage{spconf,amsmath,graphicx,hyperref}

\usepackage{booktabs}
\usepackage{multirow}

\usepackage{amssymb}
\usepackage{float}
\usepackage{caption}
\usepackage{subcaption}
\usepackage{xcolor}
\usepackage{hyperref}
\usepackage{enumitem}
\usepackage{mathtools}

\usepackage{siunitx}

\usepackage[normalem]{ulem}

\usepackage{textcomp}

\def\x{{\mathbf x}}

\newcommand {\st}{\text{\,s.t.}~}
\DeclareMathOperator{\E}{\mathbb{E}}

\usepackage[colorinlistoftodos]{todonotes}
\usepackage{balance}

\colorlet{LightTeal}{white!70!teal}
\colorlet{LightOrange}{white!70!orange}
\colorlet{LightRed}{white!70!red}

\definecolor{DarkGreen}{RGB}{1,100,32}

\usepackage{mysymbol}

\title{Graph Signal Diffusion Models for Wireless Resource Allocation}

\name{Yi\u{g}it~Berkay~Uslu, Samar~Hadou, Shirin~Saeedi~Bidokhti, Alejandro~Ribeiro}

\address{Dept. of Electrical and Systems Engineering,
University of Pennsylvania, Philadelphia, PA, USA}

\begin{document}
\ninept
\maketitle
\begin{abstract}
We consider constrained ergodic resource optimization in wireless networks with graph-structured interference. We train a diffusion model policy to match expert conditional distributions over resource allocations. By leveraging a primal-dual (expert) algorithm, we generate primal iterates that serve as draws from the corresponding expert conditionals for each training network instance. We view the allocations as stochastic graph signals supported on known channel state graphs. We implement the diffusion model architecture as a U-Net hierarchy of graph neural network (GNN) blocks, conditioned on the channel states and additional node states. At inference, the learned generative model amortizes the iterative expert policy by directly sampling allocation vectors from the near-optimal conditional distributions. In a power-control case study, we show that time-sharing the generated power allocations achieves near-optimal ergodic sum-rate utility and near-feasible ergodic minimum-rates, with strong generalization and transferability across network states.
\end{abstract}


\section{Introduction}
\label{sec:intro}

Generative models enable approximating the solutions to stochastic optimization problems whose decision variable is itself a probability distribution over high-dimensional action spaces \cite{ho2020denoising}. This is particularly relevant in physical decision-making setups, e.g., wireless systems, where optimizing quality-of-service (QoS) metrics under system requirements typically leads to constrained nonconvex problems. In such problems, whenever randomization is physically implementable, stochastic policies can attain utility-constraint trade-offs that are unattainable by any optimal deterministic policy because they induce convex combinations of deterministic operating points through time sharing \cite{neely2010stochastic}. This is especially appealing in fast-fading wireless channels, where QoS utilities and constraints are naturally evaluated in the ergodic sense \cite{naderializadeh2022stateaugmented}.

Much of the existing literature utilizes generative models either to generate domain-specific synthetic data or to perform data augmentation for learning-based wireless resource allocation
\cite{kasgari2020experienced, zhang2026improve}.
A separate line of work~\cite{meng2025conditional, diffsg2024liang, Liang2024DiffusionModelsNetworkOptimizers,  darabi2024diffusion}
proposes generative solvers for network optimization, aiming to learn solution distributions that place most of their probability mass near optimal deterministic solutions. The generative process maps random noise into high-quality solutions by denoising the samples corrupted in the forward process. In contrast to these prior works, we focus on ergodic wireless resource allocation, where the solutions are inherently stochastic.

We cast stochastic wireless resource allocation as a conditional generative modeling problem. For each network state, we learn a conditional distribution over allocation vectors matching the expert conditionals. Since the expert conditionals are unavailable in closed-form, we instead obtain approximate samples using a primal--dual expert algorithm and then train a conditional diffusion model to imitate the induced expert allocation distributions. Rather than solving a constrained optimization problem online for each new network realization, the learned diffusion model policy directly generates feasible and near-optimal allocations through conditional sampling and amortizes the iterative expert algorithm. 

We exploit the graph structure of wireless networks in our approach. Interference relationships define a graph topology with the users as nodes, and the user states are represented as signals supported on that topology. This makes graph neural networks (GNNs), which exhibit scalability and transferability \cite{MASKEY202348}, suitable parameterizations for our conditional generation task. We formulate optimal power control in ad-hoc wireless networks as a graph-signal denoising/generation problem and parameterize the diffusion policy with a U-graph neural network (U-GNN) architecture \cite{uslu2025graphsignalgenerativediffusion}.The U-GNN uses a U-shaped cascade of GNN blocks with skip connections and is conditioned jointly on channel states and user states that are represented as graphs and graph signals, respectively.

This paper extends our earlier work \cite{uslu2025generative}, which first introduced a diffusion-based generative modeling framework for the imitation learning of stochastic wireless policies. This paper expands that framework in both algorithmic detail and empirical scope. First, we provide a more complete theoretical and implementation-level treatment of the primal--dual algorithm used to construct the expert stochastic policies. Second, we study a similar power control problem over a broader family of network configurations, including varying user densities and QoS requirements. Third, we employ the U-GNN architecture as the denoising backbone of the diffusion model. Finally, we demonstrate that the learned U-GNN policies, trained on moderately-sized networks, generalize well to unseen QoS requirements and transfer to networks at larger scale.

\section{Optimal Wireless Resource Allocation}
\label{sec:expert-policy}

Consider a wireless system comprised of $N$ users (nodes). We denote by $\h \in \ccalH$ the network state.
For a given $\h$, an allocation of network resources $\x(\h) \in \reals^N$ produces a quality-of-service (QoS) utility $f_0 \big( \x(\h), \h \big) \in \reals$ that we want to maximize, along with $c$ QoS requirements collected in a vector $\bbf \big( \x(\h), \h \big) \in \reals^c$ that we must satisfy. In this paper, we study stochastic wireless resource allocation policy design problems of the form
\begin{alignat}{3} \label{problem:stochastic-functional}
    \text{P}(\h)
        \coloneqq
        \max_{\Dx} ~
             \E_{\Dx}\!\Big[f_0(\x,\h)\Big],
        ~
        \st
        ~~
        \E_{\Dx}\!\Big[\bbf(\x,\h)\Big] \geq \bb0,
\end{alignat}
where $\x(\h) \sim \Dx(\cdot \given \h)$, which is the optimization variable, denotes a conditional distribution over resource allocation variables. This abstract formulation encompasses downlink power control in ad-hoc networks (see Section~\ref{sec:experiments}), as well as analogous problems of interest in point-to-point, MIMO, and broadcast channels.

Assume that the network states are drawn from an underlying distribution $\h \sim \Dh$. We write $\Dx^\star(\cdot \given \h)$ for an optimal $\h$-conditional allocation distribution solving \eqref{problem:stochastic-functional}, and
    $\D_{\x,\h}^\star$
    =
    $\Dx^\star(\cdot \given \h)\Dh$
for the joint distribution of optimal allocations and network states. We fit a parametric family of $\h$-conditional resource allocation distributions $\Dx(\cdot \given \h;\bbtheta)$, 
with parameters $\bbtheta \in \bbTheta$, that minimizes the $\Dh$-average of conditional KL divergences,
\begin{align}\label{problem:imitation-learning}
  \bbtheta^\star
     =
           \argmin_{\bbtheta}
           \E_{\h \sim \Dh}
               \Big[
                   \kldiv[\big]{\Dx^\star(\cdot \given \h)}{\Dx(\cdot \given \h; \bbtheta)}
               \Big].
\end{align}
\noindent In \eqref{problem:imitation-learning}, we aim to learn a conditional generative policy $\Dx(\cdot \given \h;\bbtheta^\star)$ that matches the expert conditional distributions $\Dx^\star(\cdot \given \h)$ across (almost) all network states. 

Solving \eqref{problem:imitation-learning} is intractable since the expert conditionals $\Dx^\star(\cdot \given \h)$ are not available in closed-form. Instead, we assume access to samples $(\x,\h)$ drawn approximately from the joint expert distribution $\D_{\x,\h}^\star$, generated by an expert algorithm described later in Section~\ref{sec:expert-policy}. This recasts the problem as one of conditional generative modeling, where we learn a generative model $\Dx(\cdot \given \h;\bbtheta^\star)$ whose samples are distributed similarly to the expert samples  $\x \given \h \sim \Dx^\star(\cdot \given \h)$ for any $\h \sim \Dh$ given. In this paper, we instantiate $\Dx(\cdot \given \h;\bbtheta)$ as a diffusion model parametrization.

\subsection{Resource Allocation Generative Diffusion Models}

Given clean expert samples $\x_0 \given \h \sim \Dx^\star(\cdot \given \h)$, a Gaussian noising process generates corrupted samples $\x_k$ over $K$ noising steps,
\begin{align} \label{eq:ddpm:forward}
    \bbx_k = \sqrt{ \bar{\alpha}_k } \bbx_0 + \sqrt{1 - \bar{\alpha}_k } \bbepsilon, ~~~ \bbepsilon \sim \mathcal{N}(\bb0, \bbI),
\end{align}
where $\{ \bar{\alpha}_k \}_{k = 1}^{K}$ is a quantity related to the noise variance schedule. Diffusion model training learns a parametric denoiser $\bbepsilon_{\bbtheta}(\bbx_k, k; \h)$ that predicts the injected noise $\bbepsilon$ from the corrupted sample $\x_k$, conditioned on the diffusion step $k$ and the network state $\h$. This yields the standard DDPM training objective \cite{ho2020denoising},
\begin{align} \label{problem:diffusion-learning}
    \bbtheta^\star =& ~\argmin_{\bbtheta}~ \E_{\bbx_0, \h, k, \bbepsilon}  \big \| \bbepsilon_{\bbtheta} \big( \bbx_k ( \bbx_0, \bbepsilon ), k; \h \big) - \bbepsilon \big \|^2.
\end{align}
\noindent Here, the expectation is over the distribution $(\x_0, \h) \sim \Dx^*(\h)\Dh$, diffusion steps $k \sim \mathrm{Unif}(1, K)$, and noise $\bbepsilon \sim \mathcal{N}(\bb0, \bbI)$.

The learned denoiser parametrizes a reverse diffusion process that maps noisy samples back into synthetic samples distributed similarly to the clean (expert) samples. To generate samples, we use a denoising diffusion implicit model (DDIM) reverse process \cite{song2021denoising}, 
\begin{align} \label{eq:ddim-sampler}
    \x_{k-1}
    \!=\!
    \sqrt{\bar{\alpha}_{k-1}}\,\widehat{\x}_0
    \!+\!
    \sqrt{1 \!-\! \bar{\alpha}_{k-1} \!-\! \sigma_k^2}\,
    \bbepsilon_{\bbtheta^\star}(\x_k,k;\h)
    +
    \sigma_k \bbw,
\end{align}
\noindent where $\widehat{\x}_0
    =
    \big( \x_k-\sqrt{1-\bar{\alpha}_k}\,\bbepsilon_{\bbtheta^\star}(\x_k,k;\h) \big) / \sqrt{\bar{\alpha}_k}$ are denoised estimates of the clean samples, $\bbw \sim \mathcal{N}(\bb0, \bbI)$, and $\sigma_k$ controls the stochasticity of the sampler. 

At inference, we initialize $\x_K \sim \mathcal{N}(\bb0,\bbI)$ and iterate \eqref{eq:ddim-sampler} for $k=K,\ldots,1$. This sampling scheme induces a conditional generative distribution $\x_0 \sim \Dx(\cdot \cond \h; \bbtheta)$ for any given $\h$. Therefore, the generative diffusion modeling framework defined by \eqref{eq:ddpm:forward}--\eqref{eq:ddim-sampler} offers a practical, sample-based surrogate for matching $\Dx(\cdot \cond \h; \bbtheta^\star)$ to the expert conditionals $\Dx^*(\cdot \cond \h)$ in \eqref{problem:imitation-learning}. 

We particularize the denoiser $\bbepsilon_{\bbtheta}$ to a graph signal diffusion model that we discuss in Section~\ref{sec:experiments}.
Next, we describe a primal-dual algorithm that generates expert policy samples for diffusion training.

\begin{figure*}[t!]
    \centering
    \includegraphics[width=.95\linewidth]
    {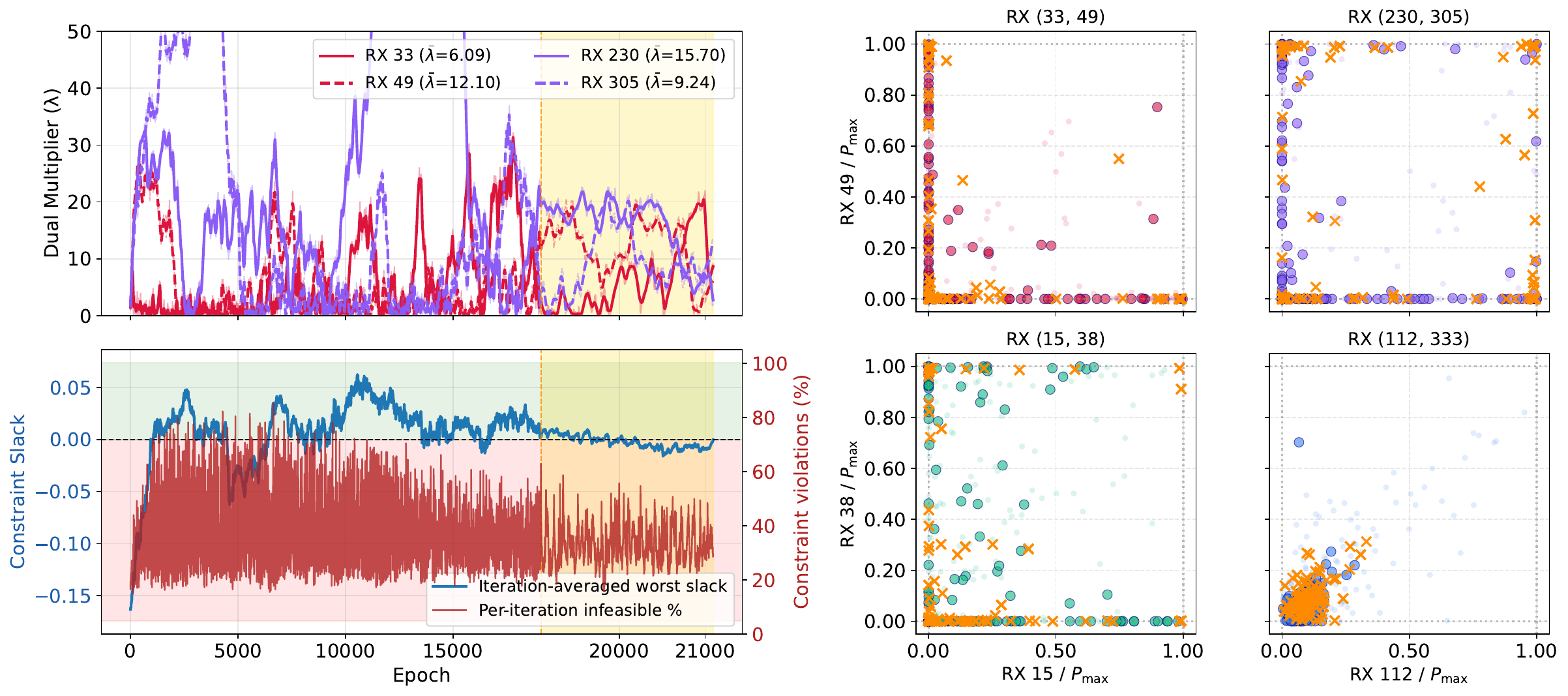}
    \vspace{-1em}
    \caption{
    \textbf{Primal-dual expert policies.} Top-left panel shows the evolution of the dual multipliers associated with the ergodic minimum-rate constraints of selected receivers for an example network $\h$ in the power-control problem of Section~\ref{sec:experiments}. The golden shaded region marks the convergence regime. Bottom-left panel plots 
    the per-iteration fraction of violated constraints against the worst-case constraint slack, averaged over a $2000$-iteration moving window. Although the iterates $\x^\dagger(\bblambda_k)$ remain infeasible for at least 20\% of the receivers, the randomized policy $\widehat{\ccalD}_{\bbx}(\cdot \cond \h)$ is near-optimal and feasible in the average-sense [cf.~\eqref{thm:near-optimality},~\eqref{thm:as-feasbility}]. The long transient preceding convergence is evident from both panels. The panels on the right show several 2D slices of the allocation vectors drawn from the expert policy in the convergence regime, with samples from the transient regime shown as faint markers. The top-row slices correspond to the receivers plotted in the top-left panel. The orange overlays show the corresponding slices for samples generated by the learned policy $\Dx(\cdot \cond \h; \bbtheta^\star)$, which closely matches the experts.
    }
    \label{fig:expert-policy}
\end{figure*}

\subsection{A Primal-Dual Expert Algorithm} \label{sec:expert-policy}

To construct the expert policies used for training, we begin with the deterministic resource-allocation problem for a fixed $\h$:

\begin{alignat}{3} \label{problem:deterministic-functional}
    \widetilde{\text{P}}(\h) 
        \coloneqq \max_{\x \in \ccalX} ~
             f_0 \big( \x(\h), \h \big), ~~ \st ~
               \bbf \big( \x(\h), \h \big) \geq \bb0.
\end{alignat}
\noindent Here, $\bbx^\star(\h)$ denotes a deterministic allocation that maximizes the utility $f_0(\x,\h)$ subject to the constraints $\bbf(\x,\h) \geq \bb0$. For given $\bblambda \succcurlyeq \bb0$, we introduce the Lagrangian relaxation of \eqref{problem:deterministic-functional},
\begin{align} \label{eq:lagrangian:deterministic-functional}
    \widetilde{\ccalL}(\x,\bblambda)
    =
    f_0(\x,\h)
    +
    \bblambda^\top \bbf(\x,\h).
\end{align}
Defining the dual function as $\widetilde g(\bblambda)
    \coloneqq
    \max_{\x \in \ccalX}
    \widetilde{\ccalL}(\x,\bblambda)$, the dual problem corresponding to \eqref{problem:deterministic-functional} is given by
\begin{align} \label{eq:dual-problem:deterministic-functional}
    \widetilde D(\h)
    =
    \min_{\bblambda \succcurlyeq \bb0}
    \widetilde g(\bblambda)
    =
    \min_{\bblambda \succcurlyeq \bb0}
    \max_{\x \in \ccalX}
    \widetilde{\ccalL}(\x,\bblambda).
\end{align}

Given a state $\h$, an initialization $\bblambda_0 \succcurlyeq \bb0$, and a dual step size $\eta_{\bblambda}>0$, a projected \emph{dual (sub)gradient descent} algorithm for \eqref{eq:dual-problem:deterministic-functional} generates the primal--dual iterates
\vspace{-5pt}
\begin{align}
    \bbx_{k+1} &= \bbx^\dagger(\bblambda_k)
    \in
    \argmax_{\bbx \in \ccalX}
    \widetilde{\ccalL}(\bbx,\bblambda_k), \label{algo:primal-update} \\
    \bblambda_{k+1}
    &=
    \Big[
        \bblambda_k
        -
        \eta_{\bblambda}\,
        \bbf(\bbx_{k+1},\h)
    \Big]_+,
    \label{algo:dual-update}
\end{align}
where $[\cdot]_+ \coloneqq \max(\cdot,0)$ denotes a componentwise projection to the nonnegative orthant. 

This algorithm alternates between a primal step that maximizes the Lagrangian for the current dual multiplier and a dual step that decreases the multipliers in the direction of positive constraint slack and increases them when constraints are violated. By parametrizing the primal variables $\x(\cdot; \bbphi)$, $\bbphi \in \bbPhi$, and approximating the exact maximization in \eqref{algo:primal-update} with multiple gradient-ascent steps on $\widetilde{\ccalL} \big( \x(\cdot; \bbphi),\bblambda_k \big)$ for given $\bblambda_k$, we obtain a \emph{primal-dual learning algorithm} as an expert policy. 

For general, nonconvex utilities $f_0$ and $\bbf$, the primal iterates of \eqref{algo:primal-update}--\eqref{algo:dual-update} need not converge to a single optimizer of \eqref{problem:deterministic-functional}. Nevertheless, the algorithm provably generates a primal trajectory that is asymptotically near-optimal and almost surely feasible \cite{naderializadeh2022stateaugmented}\footnote{The statement holds for the parametrized version, with the optimality gap additionally depending on the expressiveness of the parametrization $\bbPhi$.},
\vspace{-5pt}
\begin{align}
    \liminf_{K \to \infty}
    \Bigg[ \frac{1}{K}
    \sum_{k=1}^K
    f_0 \big( \bbx_k,\h \big) \Bigg]
    &\ge
    \text{P}(\h) - \ccalO(\eta_{\bblambda}),
    \label{thm:near-optimality} \\
    \liminf_{K \to \infty}
    \Bigg[ \frac{1}{K}
    \sum_{k=1}^K
    ~\bbf \big( \bbx_k,\h \big) \Bigg]
    &\ge
    \bb0,
    \qquad \text{a.s.}
    \label{thm:as-feasbility}
\end{align}
Note that $\text{P}(\h)$ denotes the optimum value of the stochastic problem in \eqref{problem:stochastic-functional}, and the guarantees in \eqref{thm:near-optimality}--\eqref{thm:as-feasbility} do not hold pointwise in the primal iterates, but rather in the ergodic (average) sense. Hence, for sufficiently small $\eta_{\bblambda}$, large enough $K$, and after discarding an initial burn-in window of length $K_0$, the sequence of iterates $\{\x_k\}_{k \geq K_0}^{K_0 + K}$ can be interpreted as samples from an induced stochastic policy
\vspace{-5pt}
\begin{align} \label{eq:expert-distribution}
    \widehat{\ccalD}_{\x}(\x \given \h) \coloneqq \frac{1}{K}\sum_{k=K_0}^{K_0+K} \delta(\x - \x_k),
\end{align}
where $\delta(\x - \x_k)$ denotes the Dirac measure at $\x_k$.

We illustrate the generation of primal-dual expert policy datasets in Fig.~\ref{fig:expert-policy}. Note that $\widehat{\ccalD}_{\x}(\cdot \given \h)$ approximates an expert conditional distribution $\ccalD_{\x}^\star(\cdot \given \h)$ through  a late-iterate window of the primal--dual trajectory and serves as an approximately optimal and feasible stochastic allocation policy for \eqref{problem:stochastic-functional}. It is also noteworthy that although the update rule \eqref{algo:primal-update}--\eqref{algo:dual-update} is derived from the deterministic problem \eqref{problem:deterministic-functional}, its limiting behavior is associated with the stochastic (convex) relaxation in \eqref{problem:stochastic-functional}. 
The dual descent dynamics, rather than converging to a deterministic policy, generate a \emph{stochastic time-sharing policy} whose ergodic averages attain near-optimal utility.  

\begin{remark}
Despite the availability of an expert policy with such guarantees, its direct use at inference remains computationally prohibitive because sampling from the expert policy requires running the primal-dual updates online. Even the state-augmented variants \cite{
uslu2025faststateaugmentedlearningwireless} that move the Lagrangian maximization step in \eqref{algo:primal-update} to offline training, still require online dual updates and remain subject to the trade-off governed by $\eta_{\bblambda}$ between shorter transients and improved asymptotic performance. While this trade-off can be partially mitigated by improved initialization of the dual multipliers \cite{uslu2025faststateaugmentedlearningwireless}, online dual updates remain necessary. By learning a generative model, we bypass online policy learning and eliminate the inference overhead, up to the inference cost of diffusion sampling in \eqref{eq:ddim-sampler}.
\end{remark}

\begin{figure*}[t!]
    \centering
    \includegraphics[width=\linewidth]{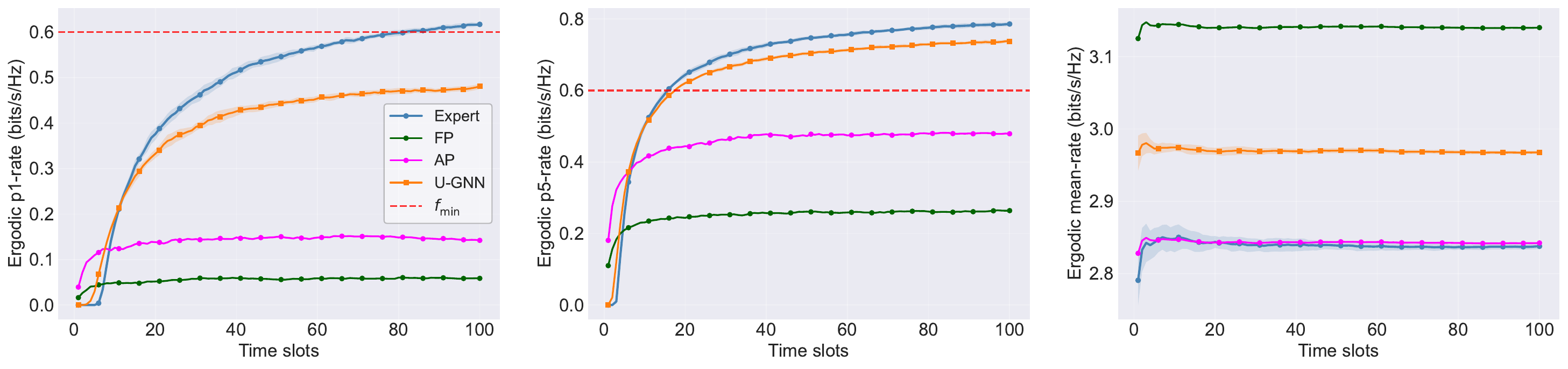}
    \vspace{-1.5em}
    \caption{\textbf{Performance of the time-shared U-GNN policy.} The U-GNN policy closely tracks the expert policy in terms of tail rates (left, middle) while achieving comparable mean-rate utility (right). The deterministic baselines fail to satisfy the QoS levels for the tail users.}
    \label{fig:test-performance:ergodic-rates}
\end{figure*}

\section{Optimal Power Control}
\label{sec:experiments}

\def \numNodes {400}
\def \numRx {\numNodes}
\def \numTestNetworksPerDensity {32}
\def \numDensities {4}
\pgfmathtruncatemacro{\numTestNetworks}{\numTestNetworksPerDensity * \numDensities}

\def \deltaTms {50}
\def \rmin {0.6}

\def \pOneExpert {0.61}   \def \pOneUgnn {0.48}
\def \pOneAP {0.14}       \def \pOneFP {0.06}
\def \pFiveExpert {0.79}  \def \pFiveUgnn {0.73}
\def \pFiveAP {0.48}      \def \pFiveFP {0.26}
\def \meanExpert {2.85}   \def \meanUgnn {2.97}
\def \meanAP {2.85}       \def \meanFP {3.14}

\def \bandwidthInMHz {40}
\def \noisePSDIndBmPerHz {-174}
\def \PmaxInMW {10}

\def \deploymentRultralow {7800}
\def \deploymentRlow {7000}
\def \deploymentRmid {6300}
\def \deploymentRhigh {5800}

\pgfmathsetmacro{\deploymentAreaultralow} {(\deploymentRultralow / 1000) * (\deploymentRultralow / 1000)}
\pgfmathsetmacro{\deploymentArealow} {(\deploymentRlow / 1000) * (\deploymentRlow / 1000)}
\pgfmathsetmacro{\deploymentAreamid} {(\deploymentRmid / 1000) * (\deploymentRmid / 1000)}
\pgfmathsetmacro{\deploymentAreahigh} {(\deploymentRhigh / 1000) * (\deploymentRhigh / 1000)}

\pgfmathsetmacro{\densityultralow} { \numNodes * (1 / \deploymentAreaultralow)}
\pgfmathsetmacro{\densitylow} { \numNodes * (1 / \deploymentArealow)}
\pgfmathsetmacro{\densitymid} { \numNodes * (1 / \deploymentAreamid)}
\pgfmathsetmacro{\densityhigh} { \numNodes * (1 / \deploymentAreahigh)}

We present an optimal power control case study that slightly modifies and extends the setups we worked with in prior works \cite{uslu2025generative, uslu2025faststateaugmentedlearningwireless}. 

We model the network state (configuration) 
$\h$ as a realization drawn from a stochastic network geometry together with a fading-channel model $\Dh$. Specifically, we adopt a random geometric graph (RGG) model for the physical network layout and deploy $N = \numNodes$ transmitter-receiver (tx-rx) pairs (users) uniformly over a square area of side length $R$ meters. Extending \cite{uslu2025generative}, we generate $32$ networks for each $R \in \{ \deploymentRhigh, \deploymentRmid, \deploymentRlow, \deploymentRultralow\}$, yielding a dataset $\ccalT = \{ \h^{(1)}, \ldots, \h^{(|\ccalT|)} \}$ of $|\ccalT| = 128$ distinct network configurations. The resulting average user densities vary as $\nu \in \{ \pgfmathprintnumber[fixed, precision=1]{\densityultralow},
\pgfmathprintnumber[fixed, precision=1]{\densitylow},
\pgfmathprintnumber[fixed, precision=1]{\densitymid},
\pgfmathprintnumber[fixed, precision=1]{\densityhigh} \}$ tx-rx pairs/\SI{}{\kilo\meter\squared}. 

The channel model includes both large-scale and small-scale Rayleigh fading. Let $h_{ij}$ denote the large-scale gain from transmitter $i$ to receiver $j$, and $h_{ij, t}$ the instantaneous gain at time slot $t$, which fluctuates due to small-scale Rayleigh fading. We collect these gains in channel state matrices $[\h]_{ij} =  h_{ij}$, and $[\h_t]_{ij} = h_{ij, t}$. The large-scale channel state is governed primarily by the path loss plus log-normal shadowing, which together capture the effects of stochastic geometry and large-scale fading. Each slot $t$ spans $\deltaTms$ milliseconds. We set the channel bandwidth and noise power spectral density (PSD) to $W = \bandwidthInMHz$ MHz and $N_0=\noisePSDIndBmPerHz$ dBm/Hz, respectively.

We optimize the transmit power levels $\x \in [0,P_{\max}]^{N}$, where $P_{\max}=\PmaxInMW$ mW is the maximum power budget. Given an allocation $\x_t$ and fading realization $\h_t$ at time slot $t$, the rate of receiver $j$ is
\begin{align} \label{eq:receiver-rate}
    \big[ \widetilde{\bbr} \big( \x_t, \h_t \big) \big]_j = \log_2 \bigg( 1 + \frac{ [\x_t]_j \cdot [\h_t]_{jj} }{W N_0 + \sum_{i\neq j} [\x_t]_i \cdot [\h_t]_{ij}}  \bigg).
\end{align}
We evaluate the policy performance by the ergodic receiver rates,
\begin{align} \label{eq:ergodic-receiver-rate}
    \bbr(\Dx, \h) = \E_{\Dx, \ccalD_{\widetilde{\h} \vert \h}} \Big[ \widetilde{\bbr}(\widetilde{\bbH}, \bbx) \Big] \approx \frac{1}{T} \sum_{t = 1}^{T} \widetilde{\bbr}(\bbx_t, \bbH_t),
\end{align}
where $\{ \bbH_t \}_{t=1}^{T}$ and $\{ \bbx_t \}_{t=1}^{T}$ are drawn independently from the stationary fast-fading process $\ccalD_{\widetilde{\h} \vert \h}$ and a power allocation policy $\Dx(\cdot \cond \h)$, respectively.
Given $\h$, the optimal power control design seeks to maximize the ergodic sum-rate utility $\mathbf{1}^\top_{N} \bbr \big( \Dx(\h), \h \big)$, subject to minimum-rate constraints $\bbr \big( \Dx(\h), \h \big) - \mathbf{1}_{N} f_{\min}$.

\subsection{U-GNN Policies}
We decompose the network state into a channel state component and a node state component. We identify the large-scale channel matrix $\bbH \in \reals^{N \times N}$ with a weighted adjacency matrix of an interference graph, where each node $j$ represents the tx-rx pair $j$, and the edge $i \to j$ encodes the channel strength from transmitter $i$ to receiver $j$. In particular, self-loops correspond to desired links, whereas off-diagonal edges correspond to interference links. 

On this graph, we define a node signal $\bbu$ with three features per node: the normalized direct-link strength, the normalized aggregate incoming interference under full-power transmission, and the QoS requirement $f_{\min}$. Thus, for each node $j=1,\ldots,N$, we set
$
[\bbu]_j = \big(h_{jj}, \sum_{i\neq j} h_{ij}, f_{\min}\big)^\top$.
The allocations $\x$ and ergodic throughput vectors $\bbr$ are likewise viewed as graph signals supported on the same graph. The U-GNN diffusion policy is then conditioned jointly on the channel matrix $\bbH$ and the node-state signal $\bbu$. Accordingly, the denoiser is written as $\bbepsilon_{\bbtheta}(\cdot,\cdot \given \bbH,\bbu)$, and the induced allocation policy takes the form $\Dx(\cdot \given \bbH,\bbu;\bbtheta)$.

\def \expertNumLayers {3}
\def \expertNumHiddenDim {64}
\def \expertMaxEpochs {3 \times 10^4}

\def \expertNumPrimalStepPerEpoch {5}
\def \expertPrimalLR {1 \times 10^{-4}}
\def \expertBatchSize {16}
\def \expertDualStepSize {0.02}
\def \expertConvergenceWindow {200}

\def \expertNumEpochsToConverge {2 \times 10^{4}}

We run separate primal--dual expert algorithms on all $|\ccalT| = \numTestNetworks$ networks, grouped by the four density levels and five QoS levels $f_{\min} \in \{ 0.4, 0.5, 0.6, 0.7, 0.8\}$ bits/s/Hz. This procedure results in 20 expert (sub)datasets spanning all density and QoS level combinations. Each expert policy is trained with early stopping, and the final $K = \expertConvergenceWindow$ primal iterates are collected to form the expert datasets. The expert policy is parameterized by a shallow, 3-layer GNN that shares the same design as the U-GNN backbone.

\def \ugnnNumGraphConvLayers {2}
\def \ugnnNumDepth {3}
\def \ugnnNumHiddenDim {64}
\def \ugnnNumCondDim {128}
\def \ugnnNumDiffusionStepDim {\ugnnNumCondDim}
\def \ugnnGraphConvHops {2}

\def \ugnnNumMaxEpochs {5000}
\def \ugnnLR {10^{-4}}
\def \ugnnNumSamplesReal {200}
\def \ugnnNumSamplesGen {100}

\def \ugnnT {500}
\def \ugnnDDIMSteps {100}

\pgfmathtruncatemacro{\numTestNetworkScenarios}{\numTestNetworks * 5}

Using these expert datasets, we train a U-GNN diffusion model, primarily based on the implementation described in \cite{uslu2025graphsignalgenerativediffusion}. The architecture has a total depth of $\ugnnNumDepth$ and stacks encoder-decoder levels in a U-shape with skip connections. The channel states $\bbH$ are processed with $\ugnnNumGraphConvLayers$-layered, $\ugnnNumHiddenDim$-channel GNN blocks with $\ugnnGraphConvHops$-hop aggregation at each encoder and decoder level, while the embeddings of diffusion step $k$ and conditional (node) features $\bbu$ both use $\ugnnNumCondDim$ channels. We employ a linear schedule of $K = \ugnnT$ noise steps and accelerated DDIM sampling with $\ugnnDDIMSteps$ steps. Overall, the dataset contains $|\ccalT| \times 5 = \numTestNetworkScenarios$ network state instances and $\ugnnNumSamplesReal$ expert samples per instance. We train U-GNN policies for a maximum of $\ugnnNumMaxEpochs$ epochs with the AdamW optimizer and a learning rate of $\ugnnLR$.  Due to limited space, we defer most implementation and training details. We will make the full code publicly available upon publication.

We compare the U-GNN and expert policies with two deterministic baselines:
    \emph{(i) Average-power transmission policy (AP)}: For each given $\h$, we fix the transmission policy to the conditional mean of the expert conditionals, i.e.,  $\bbx^{\text{AP}}_t(\h) = \E_{\Dx^\star(\h)} [\x] \approx (1/K) \sum_{k = K_0}^{K + K_0} \x_k$ at all slots [cf.~\eqref{eq:expert-distribution}].
    \emph{(ii) Full-power transmission policy (FP)}:  All transmitters use all the transmission power available, i.e., $\bbx^{\text{FP}}_t(\h) = P_{\max} \mathbf{1}_N$ at all slots $t$.

\subsection{Numerical Results}
We perform a 5:1:2 split of the dataset $\ccalT$ across training, validation, and test sets, and draw $\ugnnNumSamplesGen$ samples from the U-GNN policy for evaluating each test network. Fig.~\ref{fig:test-performance:ergodic-rates} shows the time-evolution of the cumulative ergodic p$1$-rates (1st percentile), p$5$-rates (5th percentile), and mean-rates (sum-rate utility $f_0$ normalized by $N = \numRx$) across receivers from all test networks for a reference $f_{\min} = \rmin$ scenario. 
Overall, the U-GNN policy closely tracks the primal--dual expert in the lower-tail percentiles while achieving comparable mean-rate utility. At the 5th percentile, both policies surpass the QoS threshold within 20 time-sharing slots and reach $\pFiveExpert$ and $\pFiveUgnn$~bits/s/Hz by 100 slots, respectively, with the two curves nearly superimposed over the first 30 slots and separating only mildly thereafter. The deterministic baselines, FP and AP, plateau at $\pFiveFP$ and $\pFiveAP$~bits/s/Hz and fail to meet the QoS constraint. The comparison is more nuanced at the 1st percentile, where the U-GNN reaches $\pOneUgnn$~bits/s/Hz, substantially outperforming AP at $\pOneAP$ and FP at $\pOneFP$~bits/s/Hz, but trailing the expert at $\pOneExpert$~bits/s/Hz and remaining below the QoS threshold within the 100-slot horizon. In terms of mean-rate utility, the U-GNN attains $\meanUgnn$~bits/s/Hz, above both the expert and AP at $\meanExpert$~bits/s/Hz, while FP achieves the highest aggregate throughput of $\meanFP$~bits/s/Hz at the expense of severe QoS unfairness.


The expert benchmark is run to convergence on each test network, so the comparison naturally favors the expert, and a gap emerges at the p$1$ tail. But this convergence takes roughly $\expertNumEpochsToConverge$ epochs, each comprising
$\expertNumPrimalStepPerEpoch$ primal gradient steps [cf.~Fig.~\ref{fig:expert-policy}], and this cost recurs online for every new network state. Sampling the learned policy instead takes $\ugnnDDIMSteps$
forward passes of the denoiser per allocation [cf.~\eqref{eq:ddim-sampler}], with no
backward passes and no dual updates. We also note that the optimal power control policies are generally multi-modal and stochastic [cf.~Fig.~\ref{fig:expert-policy}] as strongly interfering tx-rx pairs often alternate their transmissions to meet their requirements.

\begin{figure}[ht!]
    \centering
    \includegraphics[width=.9\linewidth]{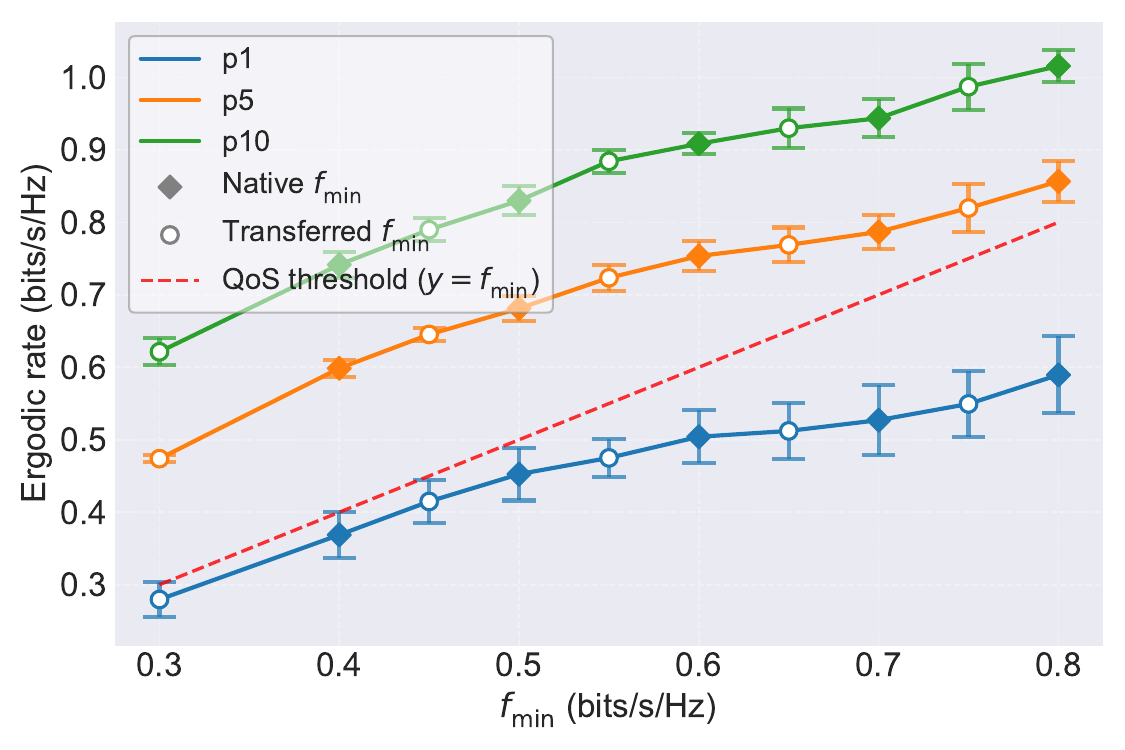}
    \vspace{-.5em}
    \caption{\textbf{Generalization of U-GNN policies across $f_{\min}$ (QoS) levels}. The model is trained on $f_{\min} \in \{0.4, 0.5, 0.6, 0.7, 0.8 \}$ and evaluated at intermediate $f_{\min}$ levels.
    }
    \label{fig:transferability:rmin}
\end{figure}

\begin{figure}[ht!]
    \centering
    \includegraphics[width=.9\linewidth]{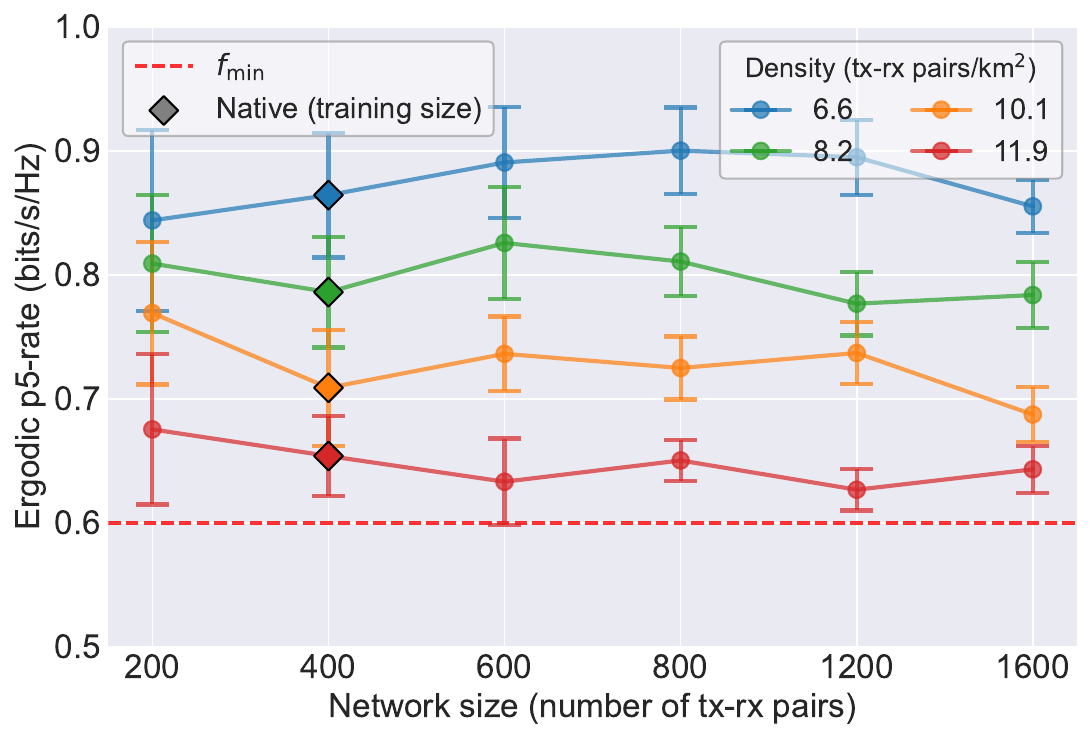}
    \vspace{-.5em}
    \caption{\textbf{Size-transferability of U-GNN policies.} The model is trained on networks of size $N = \numNodes$ and four density levels $\nu \in \{ \pgfmathprintnumber[fixed, precision=1]{\densityultralow}, \pgfmathprintnumber[fixed, precision=1]{\densitylow}, \pgfmathprintnumber[fixed, precision=1]{\densitymid}, \pgfmathprintnumber[fixed, precision=1]{\densityhigh} \}$ tx-rx pairs/\SI{}{\kilo\meter\squared}, and evaluated on  networks of varying sizes at similar densities.}
    \label{fig:transferability:size}
\end{figure}

Fig.~\ref{fig:transferability:rmin} shows the U-GNN policy generalizing across minimum-rate requirements. We plot the ergodic p$1$, p$5$, and p$10$ rates versus $f_{\min}$ on test networks pooled across all four density levels, with filled markers denoting training values and hollow markers denoting unseen values. The p$5$-rate exceeds the QoS threshold across $f_{\min} \in [0.3, 0.8]$ bits/s/Hz, confirming that at least 95\% of receivers meet the minimum-rate constraint and at least the other 4\% are near-feasible. The transferred points also interpolate smoothly, indicating that the U-GNN learns a continuous dependence on $f_{\min}$. Tail performance degrades at the out-of-distribution (OOD) endpoints, with a mild drop at low $f_{\min} = 0.3$, and sharper drops at $f_{\min} > 0.8$.

Finally, Fig.~\ref{fig:transferability:size} illustrates the size-transferability of the U-GNN policy. The model is trained on network graphs of a fixed size $|\ccalV| = \numRx$ and mixed density levels, and evaluated on test networks with varying sizes under the reference $f_{\min} = 0.6$ bits/s/Hz setting. It maintains largely stable performance across several ergodic tail-rate percentiles, indicating that the policy generalizes well across network sizes and densities. This is consistent with the U-GNN inheriting the scalability and transferability properties of its constituent GNN layers. Similar observations hold for the mean rate, additional tail percentiles, and other $f_{\min}$ levels, not shown here.

\section{Conclusion}
We proposed a generative diffusion modeling framework for stochastic resource allocation and applied it to optimal power control in ad-hoc networks. Using a primal--dual algorithm to generate expert samples and parametrizing the diffusion policy with a U-GNN architecture tailored for graph signal diffusion, we learned to sample from optimal (expert) allocation distributions for given networks (graphs) and node states. The learned policies also generalized well across QoS levels and networks of varying sizes and densities. These results show that graph signal diffusion models are promising tools for learning stochastic policies in constrained wireless optimization.


\balance
\bibliographystyle{IEEEbib}
\bibliography{strings,refs}

@article{uslu2025generative,
  title={Generative Diffusion Models for Resource Allocation in Wireless Networks},
  author={Uslu, Yigit Berkay and Hadou, Samar and Bidokhti, Shirin Saeedi and Ribeiro, Alejandro},
  journal=camsap,
  url={https/arxiv.org/abs/2504.20277},
  year={2025}
}

@article{uslu2025graphsignalgenerativediffusion,
  title={Graph Signal Generative Diffusion Models}, 
  author={Yigit Berkay Uslu and Samar Hadou and Sergio Rozada and Shirin Saeedi Bidokhti and Alejandro Ribeiro},
  journal={IEEE Intl. Conf. on Acoustics, Speech  and Signal Process. (ICASSP)},
  year={2026},
  archivePrefix={arXiv},
  primaryClass={cs.LG},
  url={https://arxiv.org/abs/2509.17250}, 
}

@article{uslu2025faststateaugmentedlearningwireless,
      title={Fast State-Augmented Learning for Wireless Resource Allocation with Dual Variable Regression}, 
      author={Yigit Berkay Uslu and Navid NaderiAlizadeh and Mark Eisen and Alejandro Ribeiro},
      journal={Under review for IEEE Transactions on Signal Processing},
      year={2026},
      eprint={2506.18748},
      archivePrefix={arXiv},
      primaryClass={eess.SP},
      url={https://arxiv.org/abs/2506.18748}, 
}

@article{naderializadeh2022stateaugmented,
  title={State-Augmented Learnable Algorithms for Resource Management in Wireless Networks},
  author={NaderiAlizadeh, Navid and Eisen, Mark and Ribeiro, Alejandro},
  journal={IEEE Transactiosn on Signal Processing},
  volume={70},
  pages={5898--5912},
  year={2022},
  publisher={IEEE}
}

@article{MASKEY202348,
title = {Transferability of graph neural networks: An extended graphon approach},
journal = {Applied and Computational Harmonic Analysis},
volume = {63},
pages = {48-83},
year = {2023},
issn = {1063-5203},
doi = {https://doi.org/10.1016/j.acha.2022.11.008},
url = {https://www.sciencedirect.com/science/article/pii/S1063520322000987},
author = {Sohir Maskey and Ron Levie and Gitta Kutyniok},
}

@article{ho2020denoising,
  title={Denoising diffusion probabilistic models},
  author={Ho, Jonathan and Jain, Ajay and Abbeel, Pieter},
  journal={Advances in neural information processing systems},
  volume={33},
  pages={6840--6851},
  year={2020}
}

@inproceedings{song2021denoising,
  title={Denoising Diffusion Implicit Models},
  year={2021},
  author={Song, Jiaming and Meng, Chenlin and Ermon, Stefano},
  booktitle=iclr
}

@book{neely2010stochastic,
  title={Stochastic network optimization with application to communication and queueing systems},
  author={Neely, Michael},
  year={2010},
  publisher={Morgan \& Claypool}
}

@article{darabi2024diffusion,
  title={Diffusion Model Based Resource Allocation Strategy in Ultra-Reliable Wireless Networked Control Systems},
  author={Darabi, Amirhassan Babazadeh and Coleri, Sinem},
  journal={IEEE Communications Letters},
  year={2024},
  publisher={IEEE}
}

@article{meng2025conditional,
  title={Conditional diffusion model with {OOD} mitigation as high-dimensional offline resource allocation planner in clustered ad hoc networks},
  author={Meng, Kechen and Zhang, Sinuo and Li, Rongpeng and Wang, Chan and Lei, Ming and Zhao, Zhifeng},
  journal={IEEE Transactions on Communications},
  volume={73},
  number={12},
  pages={14594-14609},
  year={2025},
  publisher={IEEE}
}

@ARTICLE{diffsg2024liang,
  author={Liang, Ruihuai and Yang, Bo and Yu, Zhiwen and Guo, Bin and Cao, Xuelin and Debbah, Mérouane and Poor, H. Vincent and Yuen, Chau},
  journal={IEEE Communications Magazine}, 
  title={DiffSG: A Generative Solver for Network Optimization with Diffusion Model}, 
  year={2025},
  volume={63},
  number={6},
  pages={16-24},
  doi={10.1109/MCOM.001.2400428}
}

@article{Liang2024DiffusionModelsNetworkOptimizers,
  author={Liang, Ruihuai and Yang, Bo and Chen, Pengyu and Li, Xianjin and Xue, Yifan and Yu, Zhiwen and Cao, Xuelin and Zhang, Yan and Debbah, Mérouane and Poor, H. Vincent and Yuen, Chau},
  journal={IEEE Internet of Things Journal}, 
  title={Diffusion Models as Network Optimizers: Explorations and Analysis}, 
  year={2025},
  volume={},
  number={},
  pages={1-1},
  doi={10.1109/JIOT.2025.3528955}}

@article{zhang2026improve,
  title={Improve the training efficiency of {DRL} for wireless communication resource allocation: The role of generative diffusion models},
  author={Zhang, Xinren and Yu, Jiadong},
  journal={IEEE Transactions on Wireless Communications},
  volume={25},
  pages={11593--11608},
  year={2026},
  publisher={IEEE}
}

@article{kasgari2020experienced,
  title={Experienced deep reinforcement learning with generative adversarial networks ({GANs}) for model-free ultra reliable low latency communication},
  author={Kasgari, Ali and Saad, Walid and Mozaffari, Mohammad and Poor, H. Vincent},
  journal={IEEE Transactions on Communications},
  volume={69},
  number={2},
  pages={884--899},
  year={2020},
  publisher={IEEE}
}

@string{camsap = {IEEE Intl. Workshop on Computational Advances in Multi-Sensor Adaptive Process. (CAMSAP)}}

@string{icassp = {IEEE Intl. Conf. on Acoustics, Speech  and Signal Process. (ICASSP)}}

@string{iclr = {Intl. Conf.  Learning Representations (ICLR)}}

\end{document}